\documentclass[10pt,a4paper,twoside]{article}
\usepackage{epsfig}
\usepackage{baltlat5}
\usepackage{wrapfig}
\begin{document}
\ \ \vspace{-0.5mm}

\font\dnmazasr=cmr8
\def\dnmazasit{\smalit}
\setcounter{page}{461}
\vspace{-2mm}

\titlehead{Baltic Astronomy, vol.\ts 15, 461--469, 2006.}

\titleb{A SURVEY OF COMPACT STAR CLUSTERS IN THE SOUTH-\\ WEST FIELD OF
THE M\,31 DISK.  UBVRI PHOTOMETRY}

\begin{authorl}
\authorb{D.~Narbutis}{1},
\authorb{V.~Vansevi\v{c}ius}{1},
\authorb{K.~Kodaira}{2},
\authorb{I.~\v{S}ablevi\v{c}i\={u}t\.{e}}{1},
\authorb{R.~Stonkut\.{e}}{1,3} and
\authorb{A.~Brid\v{z}ius}{1}
\end{authorl}

\begin{addressl}
\addressb{1}{Institute of Physics, Savanori\c{u} 231, Vilnius LT-02300, Lithuania \\ wladas@astro.lt}
\addressb{2}{The Graduate University for Advanced Studies (SOKENDAI), Shonan Village, Hayama, Kanagawa 240-0193, Japan}
\addressb{3}{Vilnius University Observatory, \v{C}iurlionio 29, Vilnius LT-03100, Lithuania}
\end{addressl}

\submitb{Received 2006 September 15; accepted 2006 September 30}

\begin{summary} We present the results of {\it UBVRI} broad-band
aperture CCD photometry of 51 compact star clusters located in the
South-West part of the M\,31 disk.  The mean rms errors of all measured
star cluster colors are less than 0.02 mag.  In color vs. color diagrams
the star clusters show significantly tighter sequences when compared
with the photometric data from the compiled catalog of the M\,31 star
clusters published by Galleti et al.  (2004).  \end{summary}


\resthead{Compact star clusters in the M\,31 disk}{D.~Narbutis,
V.~Vansevi\v{c}ius, K.~Kodaira et al.}

\sectionb{1}{INTRODUCTION}
Recently Kodaira et al.  (2004, hereafter Paper I) conducted a survey of
compact star clusters in the South-West part of the M\,31 disk finding
101 prominent compact objects.  In the present paper we investigate part
of these clusters using the Local Group Galaxy Survey mosaic images of
M\,31, produced by Massey et al.  (2006).  These images were used to
perform {\it UBVRI} broad-band ($R$ and $I$ bands in the Cousins system)
aperture photometry of 49 compact (KWC) and 2 compact emission (KWE)
objects from Paper I. The KWC list was supplemented by two KWE objects
satisfying the magnitude limit, $V<19$, which was generally applied for
the selection of the KWC objects.  The structural parameters of these
star clusters will be presented by \v{S}ablevi\v{c}i\={u}t\.{e} et al.
(2007, in preparation)

\sectionb{2}{DATA}
We used publicly available\footnote{~See
http://www.lowell.edu/users/massey/lgsurvey.html} stacked $U$, $B$, $V$,
$R$ and $I$ band mosaic images of M\,31 (fields F6, F7, F8, F9),
calibrated by Massey et al.  (2006), which overlap with the field
studied in Paper I. The mosaic camera used by Massey et al.  (2006)
consists of eight CCDs.  Each CCD chip covers 9\arcmin\ $\times$
18\arcmin\ field and has an individual set of color equations.  The
observations and data reductions are described in detail by Massey et
al.  (2006).

We considered mosaic images, clean from cosmic rays and cosmetically, to
be more preferable for aperture star cluster photometry than individual
exposures.  The dithering pattern of five individual exposures for each
field is the same with maximum shifts up to 1\arcmin\ in respect to the
first exposure.  Exceptions are $U$-band observations in the F9 field,
which has 6 individual exposures.  Massey et al.  (2006) do not
recommend straightforward use of their mosaic images for precise
photometry, therefore, we treat each CCD chip area on the stacked mosaic
image separately, with special care for objects residing on different
CCDs in individual exposures.

\subsectionb{2.1}{Point-spread function unification}
Mosaic images employed for aperture photometry have different
widths of stellar point-spread functions (PSF). Full width at half
maximum (FWHM) ranges from 0.7\arcsec\ to 1.3\arcsec\ (pixel size
is 0.27\arcsec). Four mosaic images have coordinate dependent PSFs
with FWHM varying more than 0.2\arcsec\ and reaching maximum of
0.3\arcsec\ for the field's F7 $I$-band mosaic image.
Consequently, this could lead to variable aperture corrections and
star cluster color bias when measured with aperture sizes as small
as $\sim$3\arcsec\ used in this study. Therefore, we applied the
DAOPHOT package of IRAF program system (Tody 1993) to compute PSF
for every mosaic image. The widest PSF (FWHM\,=\,1.3\arcsec) was
convolved with the Gaussian kernel producing the reference PSF of
FWHM\,=\,1.5\arcsec. IRAF's $psfmatch$ procedure was employed to
compute required convolution kernels for individual mosaic images
in respect to the reference PSF. The kernels were symmetrized by
replacing cores with the best fitting Gaussian profiles, and the
kernel wings with the best fitting exponential profiles truncated
at 3.5\arcsec.

IRAF's $convolve$ procedure was employed to produce mosaic images
possessing the unified PSF (FWHM\,=\,1.5\arcsec).  Finally, we achieved
uniform and coordinate-independent PSF across all convolved images.  The
maximum difference of the aperture corrections among different
photometric passband and observed field mosaic images is less than 0.02
mag.  The convolved images were photometrically calibrated and used for
star cluster photometry.

\subsectionb{2.2}{Photometric calibration}
We carefully selected well-isolated stars from Massey et al.  (2006)
Table\,4 for calibration of the {\it UBVRI} magnitudes.  The selection
criteria were as follows:  the photometric error $\sigma$\,$<$\,0.03 mag
and the number of observations $>$3 in each passband.  The total fluxes
of the calibration stars were measured by employing IRAF's $phot$
procedure through a circular aperture of 3.0\arcsec\ in diameter and by
applying the same aperture correction of 0.27 mag for all mosaic images.

Massey et al.  (2006) in Table\,2 provide color terms for individual CCD
chips of their mosaic camera and color equations.  We solved those
equations by fitting photometric zero-points for every individual CCD
chip of the mosaic images.  Typically 80 (ranging from 20 to 140)
calibration stars per chip were used.  The final errors of the derived
zero-points are less than 0.01 mag with typical fitting rms $<$0.03 mag
for the $I$-band and $<$0.02 mag for other passbands.

Color equations given by Massey et al.  (2006) supplemented with derived
zero-points were used to transform instrumental magnitudes to the
standard system.  Reliability and accuracy of calibrations were
discussed by Massey et al.  (2006) and Narbutis, Stonkut\.{e} \&
Vansevi\v{c}ius (2006).

\sectionb{3}{RESULTS}
Aperture photometry was carried out for 51 compact objects, selected
from Paper I, by employing the XGPHOT package of IRAF.  All objects are
free of visible defects on mosaic images.  Aperture centers were
determined basing on the fit of star cluster luminosity distribution
peaks on $V$-band mosaic images.  In order to avoid star cluster
photometry contamination by background objects in the crowded fields
(see Paper I, Figures 5 and 6), we decided to use individual small size
circular (for the objects KWC13, KWC20 and KWC31 -- elliptical)
apertures, indicated in Table 1. Sky backgrounds were measured in
individual circular or elliptical annuli (typical width 4\arcsec),
centered on the objects.  For some star clusters, residing on largely
variable sky background, the annuli were not centered on the objects.
These star clusters are noted by word ``sky'' in Table 1.

Our sample consists of 27 and 24 star clusters having two and three
independent measurements on different mosaic images (consequently
different CCDs), respectively.  Measured star clusters are relatively
bright objects, therefore, photon noise is virtually negligible and
photometric accuracy of the catalog is limited by the photometric
calibration procedure.  For the star clusters located in the mosaic
image areas, stacked from different CCDs, we additionally used color
equations of corresponding CCDs and performed independent
transformations to the standard system.  $V$-band magnitudes and colors,
determined in different fields (F6--F9), were averaged by assigning
smaller weights to the measurements performed on the mosaic image areas,
stacked from different CCDs.  The final catalog is provided in Table 1,
with corresponding rms errors indicated.  We artificially set the lower
accuracy limit to 0.01 mag, because of possible photometry zero-point
errors of different mosaic images.  The mean accuracy of all colors is
better than 0.02 mag.

In order to check the aperture size and centering error effects on color
accuracy we measured all objects, setting aperture sizes larger and
smaller than the standard aperture size by 10\,\%, and shifting position
of the standard aperture in both directions of the RA and Dec
coordinates by 0.15\arcsec, which corresponds to the average coordinate
discrepancy among different passband mosaic images.  Distributions of
the measured star cluster color differences were found to be consistent
with the Gaussian distribution (rms scatter $<$0.007 mag), and no
systematic bias was noticed.  Upper limit of possible color differences
due to aperture size or center variation larger than 0.015 mag (all of
them do not exceed 0.025 mag) is inherent to some colors of the clusters
KWC09, KWC15, KWC16, KWC24, KWC44, KWC46, KWC47, KWE33 and KWE52.  Their
corresponding rms errors in Table 1 are marked by asterisks.

The catalog of 51 M\,31 compact star clusters presented in Table 1 has
the following columns:

1 -- KWC or KWE number of a star cluster as defined in Paper I;

2--3 -- RA and Dec (J2000.0) are coordinates in the $V$-band mosaic
image coordinate system of field F7 (Massey et al. 2006) given in
degrees;

4 -- $V$ aperture magnitude (note that it is not the total magnitude);

5--8 -- $U$$-$$B$, $B$$-$$V$, $V$$-$$R$, $R$$-$$I$ colors with
corresponding rms errors in the line below;

9 -- the number of independent measurements ($n$) in different fields;

10 -- diameter of circular aperture or major axis length of elliptic
aperture (Ap) in arcseconds;

\vbox{
\centerline{\psfig{figure=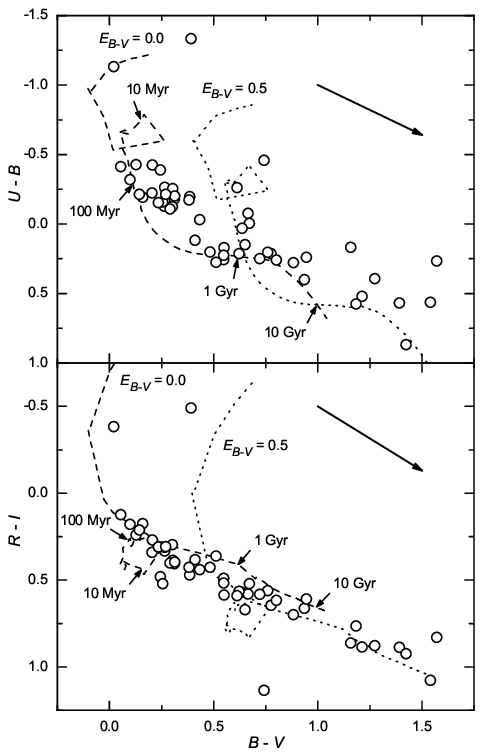,width=108mm,angle=0,clip=}}
\vspace{.5mm}
\captionb{1}{Color-color diagrams of the 51 star
clusters (open circles; Table\,1). P\'{E}GASE SSP models
($Z=0.02$, age range from 1 Myr to 15 Gyr) with $E_{B-V}=0$
(dashes) and $E_{B-V}=0.5$ (dots) are over-plotted. The reddening
vectors, corresponding to the standard Milky Way extinction law,
are indicated. Small arrows indicate the SSP ages of 10 Myr,
100 Myr, 1 Gyr and 10 Gyr.}}
\newpage

\vbox{
\centerline{\psfig{figure=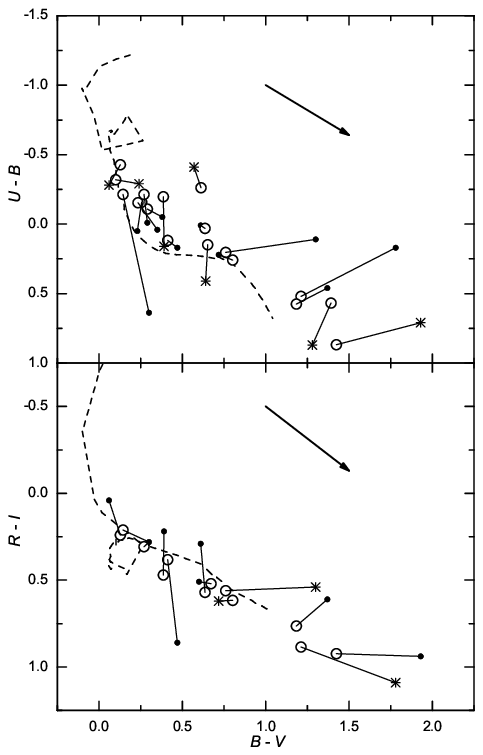,width=108mm,angle=0,clip=}}
\vspace{.5mm}
\captionb{2}{Color-color diagrams of the
corresponding star clusters taken from the catalog presented in
Table\,1 (open circles) and from Galleti et al. (2004) (dots -- CCD
photometry; asterisks -- photographic photometry).
Lines connect the same object in both catalogs. SSP models and
reddening vectors are the same as in Fig. 1. In total 18 and 12
star clusters are shown in the top and bottom panels,
respectively.}}
\newpage

\begin{center}
\vbox{\small
\tabcolsep=2.5pt
\begin{tabular}{cccrrrrrccc}
\multicolumn{11}{c}{\parbox{120mm}{{\normbf Table 1.}
{\norm {\it UVBRI} photometry catalog of the M\,31 compact star
clusters. }}}\\
\noalign{\smallskip}
\tablerule
\noalign{\smallskip}
\multicolumn{1}{c}{Cluster} &
\multicolumn{1}{c}{RA(2000)} &
\multicolumn{1}{c}{Dec(2000)} &
\multicolumn{1}{r}{\dnmazasit V~~~} &
\multicolumn{1}{r}{\dnmazasit U\,--\,B} &
\multicolumn{1}{r}{\dnmazasit B\,--\,V} &
\multicolumn{1}{r}{\dnmazasit V\,--\,R} &
\multicolumn{1}{r}{\dnmazasit R\,--\,I}~ &
\multicolumn{1}{c}{$n$} &
\multicolumn{1}{c}{Ap} &
\multicolumn{1}{c}{Note} \\
\tablerule
\noalign{\smallskip}
                    KWC01 & 10.04577 & 40.60326 &  18.498  &~--0.426  &~  0.206  &~  0.223  &~  0.269  & 3 & 3.0 & ...  \\
                          &          &          &   0.012  &   0.020  &   0.010  &   0.010  &   0.010  &   &     &      \\
                    KWC02 & 10.05932 & 40.65581 &  18.824  & --0.265  &   0.265  &   0.227  &   0.332  & 3 & 3.6 & ...  \\
                          &          &          &   0.024  &   0.010  &   0.018  &   0.011  &   0.013  &   &     &      \\
                    KWC03 & 10.06074 & 40.62242 &  18.566  & --0.413  &   0.054  &   0.096  &   0.124  & 3 & 3.0 & ...  \\
                          &          &          &   0.010  &   0.023  &   0.011  &   0.010  &   0.010  &   &     &      \\
                    KWC04 & 10.06408 & 40.61515 &  18.590  & --0.180  &   0.312  &   0.262  &   0.406  & 3 & 3.2 & ...  \\
                          &          &          &   0.025  &   0.024  &   0.016  &   0.011  &   0.010  &   &     &      \\
                    KWC05 & 10.06460 & 40.66653 &  18.486  & --0.254  &   0.303  &   0.284  &   0.387  & 3 & 3.4 & ...  \\
                          &          &          &   0.010  &   0.022  &   0.010  &   0.010  &   0.010  &   &     &      \\
                    KWC06 & 10.07201 & 40.65136 &  18.125  & --0.428  &   0.127  &   0.155  &   0.241  & 3 & 3.6 & ...  \\
                          &          &          &   0.010  &   0.013  &   0.010  &   0.010  &   0.010  &   &     &      \\
                    KWC07 & 10.07320 & 40.65556 &  18.560  & --0.320  &   0.099  &   0.112  &   0.179  & 3 & 3.0 & ...  \\
                          &          &          &   0.010  &   0.027  &   0.015  &   0.010  &   0.010  &   &     &      \\
                    KWC08 & 10.07620 & 40.54577 &  18.559  &   0.116  &   0.411  &   0.293  &   0.383  & 3 & 3.6 & ...  \\
                          &          &          &   0.014  &   0.010  &   0.010  &   0.010  &   0.018  &   &     &      \\
                    KWC09 & 10.07855 & 40.52101 &  19.329  &   0.266  &   1.570  &   0.770  &   0.830  & 2 & 3.0 & ...  \\
                          &          &          &   0.022  &   0.035\rlap{*} &   0.014\rlap{*} &   0.011  &   0.024  &   &     &      \\
                    KWC10 & 10.08099 & 40.62479 &  18.687  & --0.197  &   0.385  &   0.332  &   0.470  & 3 & 3.0 & ...  \\
                          &          &          &   0.010  &   0.013  &   0.010  &   0.010  &   0.010  &   &     &      \\
                    KWC11 & 10.08302 & 40.51324 &  18.742  & --0.193  &   0.159  &   0.142  &   0.175  & 2 & 3.4 & ...  \\
                          &          &          &   0.021  &   0.010  &   0.010  &   0.010  &   0.010  &   &     &      \\
                    KWC12 & 10.09631 & 40.51323 &  17.477  & --0.005  &   0.671  &   0.465  &   0.521  & 2 & 6.8 & ...  \\
                          &          &          &   0.017  &   0.010  &   0.013  &   0.010  &   0.020  &   &     &      \\
                    KWC13 & 10.10304 & 40.63052 &  17.221  &   0.569  &   1.391  &   0.809  &   0.888  & 3 & 6.0 & 0.55, 50  \\
                          &          &          &   0.010  &   0.017  &   0.017  &   0.010  &   0.015  &   &     &      \\
                    KWC14 & 10.10351 & 40.81297 &  19.227  &   0.171  &   0.550  &   0.440  &   0.585  & 2 & 3.0 & sky  \\
                          &          &          &   0.010  &   0.010  &   0.020  &   0.010  &   0.014  &   &     &      \\
                    KWC15 & 10.10370 & 40.81690 &  19.693  &   0.213  &   0.776  &   0.591  &   0.645  & 2 & 3.0 & sky  \\
                          &          &          &   0.010  &   0.017\rlap{*} &   0.024  &   0.011  &   0.012  &   &     &      \\
                    KWC16 & 10.10807 & 40.62813 &  19.326  &   0.168  &   1.158  &   0.778  &   0.862  & 3 & 3.0 & ...  \\
                          &          &          &   0.010  &   0.021\rlap{*} &   0.023\rlap{*} &   0.010  &   0.016  &   &     &      \\
                    KWC17 & 10.11374 & 40.75674 &  18.740  & --0.224  &   0.205  &   0.238  &   0.341  & 2 & 3.0 & ...  \\
                          &          &          &   0.014  &   0.010  &   0.011  &   0.010  &   0.018  &   &     &      \\
                    KWC18 & 10.13581 & 40.83711 &  19.140  & --0.127  &   0.264  &   0.230  &   0.311  & 2 & 3.0 & ...  \\
                          &          &          &   0.010  &   0.010  &   0.010  &   0.010  &   0.010  &   &     &      \\
                    KWC19 & 10.15227 & 40.67085 &  19.027  &   0.030  &   0.636  &   0.460  &   0.571  & 3 & 3.0 & ...  \\
                          &          &          &   0.016  &   0.010  &   0.010  &   0.010  &   0.010  &   &     &      \\
                    KWC20 & 10.15519 & 40.65408 &  19.144  & --0.077  &   0.665  &   0.455  &   0.580  & 3 & 3.4 & 0.75, 120  \\
                          &          &          &   0.019  &   0.014  &   0.010  &   0.010  &   0.014  &   &     &      \\
                    KWC21 & 10.15569 & 40.81267 &  19.255  &~--0.126  &~  0.302  &~  0.223  &~  0.297  & 2 & 3.0 & ...  \\
                          &          &          &   0.010  &   0.013  &   0.010  &   0.010  &   0.010  &   &     &      \\
                    KWC22 & 10.17370 & 40.64113 &  18.726  &   0.521  &   1.212  &   0.758  &   0.885  & 3 & 3.4 & ...  \\
                          &          &          &   0.011  &   0.013  &   0.010  &   0.015  &   0.014  &   &     &      \\
\tablerule
\end{tabular}
}
\end{center}

\begin{center}
\vbox{\small
\tabcolsep=2.5pt
\begin{tabular}{cccrrrrrccc}
\multicolumn{11}{c}{\parbox{120mm}{{\normbf Table 1.}
{\norm Continued}}}\\
\noalign{\smallskip}
\tablerule
\noalign{\smallskip}
\multicolumn{1}{c}{Cluster} &
\multicolumn{1}{c}{RA(2000)} &
\multicolumn{1}{c}{Dec(2000)} &
\multicolumn{1}{r}{\dnmazasit V~~~} &
\multicolumn{1}{r}{\dnmazasit U\,--\,B} &
\multicolumn{1}{r}{\dnmazasit B\,--\,V} &
\multicolumn{1}{r}{\dnmazasit V\,--\,R} &
\multicolumn{1}{r}{\dnmazasit R\,--\,I}~ &
\multicolumn{1}{c}{$n$} &
\multicolumn{1}{c}{Ap} &
\multicolumn{1}{c}{Note} \\
\tablerule
\noalign{\smallskip}
                    KWC23 & 10.17619 & 40.60127 &  19.226  &   0.203  &   0.759  &   0.469  &   0.560  & 3 & 3.0 & ...  \\
                          &          &          &   0.010  &   0.010  &   0.010  &   0.013  &   0.024  &   &     &      \\
                    KWC24 & 10.18421 & 40.74610 &  19.815  &   0.563  &   1.541  &   1.014  &   1.078  & 2 & 3.0 & ...  \\
                          &          &          &   0.010  &   0.017\rlap{*} &   0.021\rlap{*} &   0.010  &   0.010  &   &     &      \\
                    KWC25 & 10.19348 & 40.86130 &  19.182  & --0.174  &   0.382  &   0.316  &   0.426  & 2 & 3.2 & ...  \\
                          &          &          &   0.010  &   0.010  &   0.010  &   0.011  &   0.015  &   &     &      \\
                    KWC26 & 10.20151 & 40.58503 &  18.406  & --0.108  &   0.291  &   0.252  &   0.401  & 3 & 3.0 & ...  \\
                          &          &          &   0.010  &   0.010  &   0.010  &   0.012  &   0.011  &   &     &      \\
                    KWC27 & 10.20159 & 40.86619 &  18.620  & --0.388  &   0.245  &   0.316  &   0.480  & 3 & 3.0 & ...  \\
                          &          &          &   0.012  &   0.025  &   0.010  &   0.010  &   0.010  &   &     &      \\
                    KWC28 & 10.20241 & 40.96009 &  19.280  &   0.257  &   0.548  &   0.396  &   0.490  & 2 & 3.0 & sky  \\
                          &          &          &   0.010  &   0.020  &   0.010  &   0.010  &   0.014  &   &     &      \\
                    KWC29 & 10.20591 & 40.69223 &  18.283  &   0.258  &   0.801  &   0.501  &   0.617  & 2 & 3.4 & ...  \\
                          &          &          &   0.010  &   0.010  &   0.010  &   0.010  &   0.010  &   &     &      \\
                    KWC30 & 10.21459 & 40.55770 &  19.273  &   0.148  &   0.651  &   0.450  &   0.670  & 3 & 3.2 & ...  \\
                          &          &          &   0.010  &   0.022  &   0.010  &   0.013  &   0.021  &   &     &      \\
                    KWC31 & 10.21512 & 40.73504 &  17.982  & --0.458  &   0.742  &   0.670  &   1.135  & 2 & 4.2 & 0.70, 30  \\
                          &          &          &   0.010  &   0.014  &   0.010  &   0.010  &   0.014  &   &     &      \\
                    KWC32 & 10.21786 & 40.89896 &  19.287  &   0.401  &   0.936  &   0.568  &   0.662  & 3 & 3.2 & ...  \\
                          &          &          &   0.010  &   0.022  &   0.011  &   0.010  &   0.012  &   &     &      \\
                    KWC33 & 10.21782 & 40.97817 &  19.450  &   0.213  &   0.622  &   0.447  &   0.565  & 2 & 3.0 & sky  \\
                          &          &          &   0.010  &   0.020  &   0.010  &   0.010  &   0.011  &   &     &      \\
                    KWC34 & 10.22069 & 40.58883 &  18.878  &   0.277  &   0.882  &   0.576  &   0.699  & 3 & 4.4 & ...  \\
                          &          &          &   0.010  &   0.016  &   0.014  &   0.010  &   0.011  &   &     &      \\
                    KWC35 & 10.23960 & 40.74087 &  19.079  &   0.393  &   1.273  &   0.810  &   0.877  & 2 & 3.0 & ...  \\
                          &          &          &   0.010  &   0.010  &   0.010  &   0.010  &   0.018  &   &     &      \\
                    KWC36 & 10.27868 & 40.57474 &  19.146  & --0.201  &   0.313  &   0.288  &   0.396  & 3 & 3.0 & ...  \\
                          &          &          &   0.011  &   0.021  &   0.020  &   0.010  &   0.022  &   &     &      \\
                    KWC37 & 10.28315 & 40.88356 &  18.700  &   0.239  &   0.945  &   0.584  &   0.608  & 3 & 3.0 & ...  \\
                          &          &          &   0.017  &   0.012  &   0.010  &   0.012  &   0.021  &   &     &      \\
                    KWC38 & 10.29172 & 40.96973 &  19.013  &   0.249  &   0.721  &   0.461  &   0.582  & 2 & 3.0 & ...  \\
                          &          &          &   0.010  &   0.017  &   0.010  &   0.010  &   0.017  &   &     &      \\
                    KWC39 & 10.30333 & 40.57155 &  18.104  & --0.158  &   0.256  &   0.266  &   0.521  & 2 & 3.6 & sky  \\
                          &          &          &   0.010  &   0.010  &   0.010  &   0.010  &   0.015  &   &     &      \\
                    KWC40 & 10.30768 & 40.56615 &  18.321  & --0.155  &   0.234  &   0.216  &   0.310  & 2 & 3.6 & ...  \\
                          &          &          &   0.010  &   0.015  &   0.010  &   0.010  &   0.019  &   &     &      \\
                    KWC41 & 10.32568 & 40.73369 &  19.181  &~  0.225  &~  0.549  &~  0.362  &~  0.517  & 2 & 3.2 & ...  \\
                          &          &          &   0.010  &   0.010  &   0.010  &   0.010  &   0.010  &   &     &      \\
                    KWC42 & 10.32810 & 40.95436 &  17.666  &   0.869  &   1.423  &   0.830  &   0.923  & 2 & 4.0 & ...  \\
                          &          &          &   0.010  &   0.016  &   0.010  &   0.010  &   0.022  &   &     &      \\
                    KWC43 & 10.33723 & 40.98452 &  18.114  &   0.575  &   1.184  &   0.685  &   0.764  & 2 & 3.4 & ...  \\
                          &          &          &   0.010  &   0.023  &   0.010  &   0.010  &   0.010  &   &     &      \\
                    KWC44 & 10.35044 & 40.61307 &  18.687  & --0.213  &   0.144  &   0.145  &   0.212  & 2 & 4.0 & ...  \\
                          &          &          &   0.010  &   0.010  &   0.010  &   0.010  &   0.027\rlap{*} &   &     &      \\
\tablerule
\end{tabular}
}
\end{center}

\begin{center}
\vbox{\small
\tabcolsep=3.5pt
\begin{tabular}{cccrrrrrccc}
\multicolumn{11}{c}{\parbox{120mm}{{\normbf Table 1.}
{\norm Continued}}}\\
\noalign{\smallskip}
\tablerule
\noalign{\smallskip}
\multicolumn{1}{c}{Cluster} &
\multicolumn{1}{c}{RA(2000)} &
\multicolumn{1}{c}{Dec(2000)} &
\multicolumn{1}{r}{\dnmazasit V~~~} &
\multicolumn{1}{r}{\dnmazasit U\,--\,B} &
\multicolumn{1}{r}{\dnmazasit B\,--\,V} &
\multicolumn{1}{r}{\dnmazasit V\,--\,R} &
\multicolumn{1}{r}{\dnmazasit R\,--\,I}~ &
\multicolumn{1}{c}{$n$} &
\multicolumn{1}{c}{Ap} &
\multicolumn{1}{c}{Note} \\
\tablerule
\noalign{\smallskip}
                    KWC45 & 10.36251 & 40.69373 &  19.221  & --0.032  &   0.433  &   0.341  &   0.440  & 2 & 3.0 & ...  \\
                          &          &          &   0.010  &   0.010  &   0.029  &   0.010  &   0.026  &   &     &      \\
                    KWC46 & 10.40363 & 40.79040 &  18.691  &   0.201  &   0.483  &   0.334  &   0.426  & 3 & 3.0 & ...  \\
                          &          &          &   0.014  &   0.018  &   0.017  &   0.010  &   0.014\rlap{*} &   &     &      \\
                    KWC47 & 10.40855 & 40.56967 &  18.948  & --0.263  &   0.612  &   0.393  &   0.590  & 2 & 3.0 & ...  \\
                          &          &          &   0.010  &   0.021\rlap{*} &   0.010  &   0.017  &   0.010  &   &     &      \\
                    KWC48 & 10.41051 & 40.82676 &  19.293  &   0.276  &   0.512  &   0.327  &   0.362  & 3 & 3.0 & ...  \\
                          &          &          &   0.017  &   0.022  &   0.011  &   0.010  &   0.011  &   &     &      \\
                    KWC49 & 10.41184 & 40.68182 &  18.111  & --0.214  &   0.270  &   0.238  &   0.308  & 2 & 3.0 & ...  \\
                          &          &          &   0.010  &   0.028  &   0.010  &   0.010  &   0.010  &   &     &      \\
                    KWE33 & 10.19922 & 40.98502 &  18.459  & --1.335  &   0.392  &   0.294  & --0.491  & 2 & 4.0 & ...  \\
                          &          &          &   0.010  &   0.031\rlap{*} &   0.018  &   0.010  &   0.010\rlap{*} &   &     &      \\
                    KWE52 & 10.41127 & 40.73314 &  18.793  & --1.132  &   0.021  &   0.502  & --0.384  & 2 & 3.6 & ...  \\
                          &          &          &   0.011  &   0.026\rlap{*} &   0.010  &   0.010\rlap{*} &   0.011\rlap{*} &   &     &      \\
\tablerule
\end{tabular}
}
\end{center}

\vskip5mm
11 -- notes (for elliptic apertures:  ratio of minor to major axis and
position angle of the major axis, calculated counterclockwise from the
North direction; ``sky'' - sky background values measured in the annulus
not centered on the objects).

Color-color diagrams of the compact star clusters from Table\,1 are
shown in Figure 1 with over-plotted simple stellar population (SSP)
models of solar metallicity ($Z=0.02$) computed with P\'{E}GASE (Fioc \&
Rocca-Volmerange 1997).  Default P\'{E}GASE parameters and universal
initial mass function (IMF) by Kroupa (2002) were applied.  Reddening
arrows are depicted by applying the color excess ratios of the standard
extinction law ($V$-band extinction to color excess ratio,
$A_{V}/E_{B-V}=3.1$).  Galactic interstellar extinction in the direction
of the South-West part of the M\,31 galaxy was estimated by employing
NASA Extragalactic Database Galactic Reddening and Extinction
Calculator, $E_{B-V}=0.06$.  However, star cluster photometry data,
plotted in Figures 1 and 2, are not dereddened.

We compared our catalog data with the data from the compiled catalog
published by Galleti et al.  (2004; version V.2.0, May
2006)\footnote{~See http://www.bo.astro.it/M31/}.  In total 27 star
clusters were cross-identified (18 and 12 objects have $U$$-$$B$ and
$R$$-$$I$ colors, respectively) and are shown in Figure 2. Our
photometric data exhibit significantly smaller scatter in color-color
diagrams, which is a result of homogeneous photometric survey of the
M\,31 galaxy performed by Massey et al.  (2006) and individual apertures
carefully chosen for each star cluster.  Therefore, we conclude that
accuracy of the measured aperture colors of the M\,31 compact star
clusters is satisfactory to be used for star cluster parameter (age,
extinction, metallicity) determination basing on comparison with SSP
models.  Photometry data analysis of the measured compact clusters will
be presented elsewhere (Vansevi\v{c}ius et al. 2007, in preparation).

\vskip5mm

ACKNOWLEDGMENTS.  This work was financially supported in part by a Grant
T-08/06 of the Lithuanian State Science and Studies Foundation.  This
research has made use of the NASA/IPAC Extragalactic Database (NED)
which is operated by the Jet Propulsion Laboratory, California Institute
of Technology, under contract with the National Aeronautics and Space
Administration, and of SAOImage DS9, developed by Smithsonian
Astrophysical Observatory. We are thankful to Valdas Vansevi\v cius
for correcting the manuscript.

\vskip5mm

\References
\refb Fioc~M., Rocca-Volmerange~B. 1997, A\&A, 326, 950

\refb Galleti~S., Federici~L., Bellazzini~M., Fusi Pecci~F., Macrina~S.
2004, A\&A, 416, 917

\refb Kodaira~K., Vansevi\v{c}ius~V., Brid\v{z}ius~A., Komiyama~Y.,
Miyazaki~S., Stonkut\.{e}~R., \v{S}ablevi\v{c}i\={u}t\.{e}~I.,
Narbutis~D. 2004, PASJ, 56, 1025

\refb Kroupa~P. 2002, Science, 295, 82

\refb Massey~P., Olsen~K.\,A.\,G., Hodge~P.  W., Strong~S.  B.,
Jacoby~G.  H., Schlingman~W., Smith~R.  C. 2006, AJ, 131, 2478

\refb Narbutis~D., Stonkut\.{e}~R., Vansevi\v{c}ius~V. 2006, Baltic
Astronomy, 15, 471 (this issue)

\refb Tody~D. 1993, in {\it Astronomical Data Analysis Software and
Systems II}, eds.  R. J.~Hanisch, R.\,J.\,V.~Brissenden \& J.~Barnes,
ASP Conf.  Ser., 52, 173

\end{document}